\documentclass[aps, reprint,pra,nofootinbib,longbibliography]{revtex4-1}
\usepackage{dcolumn}
\usepackage[utf8]{inputenc}
\usepackage{amsmath}
\usepackage{amsfonts}
\usepackage{amssymb}
\usepackage{graphicx}
\usepackage{hyperref}
\usepackage{xcolor}

\newcommand{\ket}[1]{\ensuremath{\left|{#1}\right\rangle}}
\newcommand{\bra}[1]{\ensuremath{\left\langle{#1}\right |}}

\newcommand{\mb}{\mathbf}
\newcommand{\bs}{\boldsymbol}

\begin{document}

\title{Gauge invariance of the Aharonov-Bohm effect in a quantum electrodynamics framework}

\author{Pablo L. Saldanha}\email{saldanha@fisica.ufmg.br}
\affiliation{Departamento de F\'isica, Universidade Federal de Minas Gerais, Belo Horizonte, MG 31270-901, Brazil}

\date{\today}

\begin{abstract}
The gauge invariance of the Aharonov-Bohm (AB) effect with a quantum treatment for the electromagnetic field is demonstrated. We provide an exact solution for the electromagnetic ground energy due to the interaction of the quantum electromagnetic field with the classical charges and currents that act as sources of the potentials in a classical description, in the Lorenz gauge. Then, we use first-order perturbation theory to compute an extra change on the electromagnetic ground energy due to the presence of a quantum charged particle with known wave function in the system. This energy in general depends on the quantum particle path in an interferometer, what results in an AB phase difference between the paths. The gauge invariance of this AB phase difference is then shown for the magnetic, electric, and the recently proposed electrodynamic versions of the AB effect. However, the AB phase difference could depend on the gauge for nonclosed paths, what reinforces the view that it only can be measured in closed paths.
\end{abstract}


\maketitle

\section{Introduction}

The Aharonov-Bohm (AB) effect \cite{ehrenberg49,aharonov59} has a fundamental importance in Physics, since it brings important questions regarding the locality of the electromagnetic interactions. In the magnetic AB effect \cite{ehrenberg49,aharonov59,chambers60,tonomura86}, a quantum charged particle has two possible paths in an interferometer that enclose a magnetic flux. The interference pattern depends on the enclosed flux even if the particle propagates in regions where the magnetic field is null, but the vector potential is not, with the appearance of an AB phase difference between the paths. In the electric AB effect \cite{aharonov59,matteucci85,oudenaarden98}, the interference pattern depends on the included scalar potential difference between the paths, even if the particle only propagates in regions with a null electric field. Recently, an electrodynamic AB effect was proposed \cite{saldanha23}, with the current in a solenoid outside the interferometer varying while the quantum particle is in a superposition state inside two Faraday cages. Somewhat surprisingly, a nonzero AB phase difference appears even if the particle paths enclose no magnetic flux and are subjected to no scalar potential difference \cite{saldanha23}. In all these situations, a description of the phenomenon in terms of a local interaction of the quantum charged particle with the classical electromagnetic field is not possible. We can describe it through a local interaction of the particle with the electromagnetic potentials \cite{aharonov59} or through a nonlocal interaction of the particle with the electromagnetic fields \cite{peshkin81,kang13,saldanha16,pearle17a,li22}.

To address locality issues in the AB effect, Santos and Gonzalo described the magnetic AB effect in an interferometer with cylindrical symmetry using a quantum electromagnetic field to mediate the interactions \cite{santos99}. Recently, Marletto and Vedral discussed how such treatment can justify a local acquirement of the AB phase by the quantum particle while it propagates through the interferometer \cite{marletto20}. In posterior works, the treatment was improved using the quantum electrodynamics formalism in the Lorenz gauge \cite{saldanha21a,saldanha21b}, permitting the treatment of the magnetic AB effect for arbitrary interferometer geometries, the electric AB effect, the Aharonov-Casher effect \cite{aharonov84}, and the magnetic AB effect with shielding \cite{tonomura86}. Using second-order perturbation theory, these works show that the energy of the ground state of the quantum electromagnetic field depends on the particle path, this behavior being the responsible for the appearance of the AB phase \cite{santos99,marletto20,saldanha21a,saldanha21b}. This dependence can be associated to an exchange of photons between the quantum particle and the sources of the potentials, in a local description of the phenomenon.

An important question regards the gauge invariance of the AB effect in a description using a quantum electromagnetic field. Kang showed this gauge invariance treating the magnetic AB effect in a simplified two-dimension case with a charged particle and a fluxon \cite{kang22}. Hayashi, on the other hand, showed that the electromagnetic ground energy changes in different ways for two particular gauges in the magnetic AB effect \cite{hayashi23}, criticizing previous works that supposedly propose ways to measure the AB phase in nonclosed paths \cite{marletto20,saldanha21a}. Nonetheless, the predicted AB phases coincide for closed paths \cite{hayashi23}. 

Here we demonstrate the gauge invariance of the electric, magnetic, and electrodynamic versions of the AB effect for general gauges in a quantum electrodynamics treatment. We provide an exact solution for the electromagnetic ground energy considering the interaction of the quantum electromagnetic field with the system classical charges and currents (which are responsible for producing the potentials in a classical description) in the Lorenz gauge. Using first-order perturbation theory, we compute an extra change on this ground energy due to the interaction of the quantum electromagnetic field with a quantum charged particle with known wave function. The obtained ground energy agrees with previous treatments using second-order perturbation theory the Lorenz gauge \cite{saldanha21a,saldanha21b}. The dependence of this ground energy on the quantum particle path in an interferometer results in an AB phase difference between the paths. By performing gauge changes on the potential operators of the quantum electromagnetic field, we demonstrate the gauge invariance of the AB effect. As Hayashi \cite{hayashi23}, we conclude that the gauge invariance exists only for closed paths, such that a measurement of the AB phase in nonclosed paths, as recently proposed \cite{marletto20}, should not be possible. But we discuss that the proposal of Ref. \cite{saldanha21a}, on the other hand, is not for measuring the AB phase in a nonclosed path, but for measuring the AB phase in a closed path with varying electromagnetic fields. We discuss that the intermediate AB phase predicted in this work \cite{saldanha21a}, related to the electrodynamic AB effect \cite{saldanha23}, is indeed gauge-invariant.

\section{AB schemes with a classical electromagnetic field}

Fig. 1 displays a setup that could be used to demonstrate different kinds of AB effects for quantum non-relativistic charged particles. It consists of a Mach-Zehnder interferometer with an ``infinite'' solenoid at its interior and two Faraday cages, one in each path. Let us first describe it when the solenoid and the Faraday cages are absent (or when the solenoid has a null magnetic field and the cages are in a null scalar potential). The beam splitter BS$_1$ divides the incident wave function into components $\Psi_a^0(\mb{r},t)$ and $\Psi_b^0(\mb{r},t)$ that propagate through the different paths and recombine at the beam splitter BS$_2$, resulting in interference. For simplicity, consider that the wave functions $\Psi_a^0(\mb{r},t)$ and $\Psi_b^0(\mb{r},t)$ are well localized in space around central positions $\mb{r}_a(t)$ and $\mb{r}_b(t)$ respectively, such that the potentials to be included do not vary much around the regions where these functions have non-negligible values. Let us also consider that the momenta of each wave packet are reasonably well defined, having average values $\mb{p}_a(t)$ and $\mb{p}_b(t)$, but of course respecting the uncertainty relations. 

\begin{figure}
  \centering
    \includegraphics[width=8.5cm]{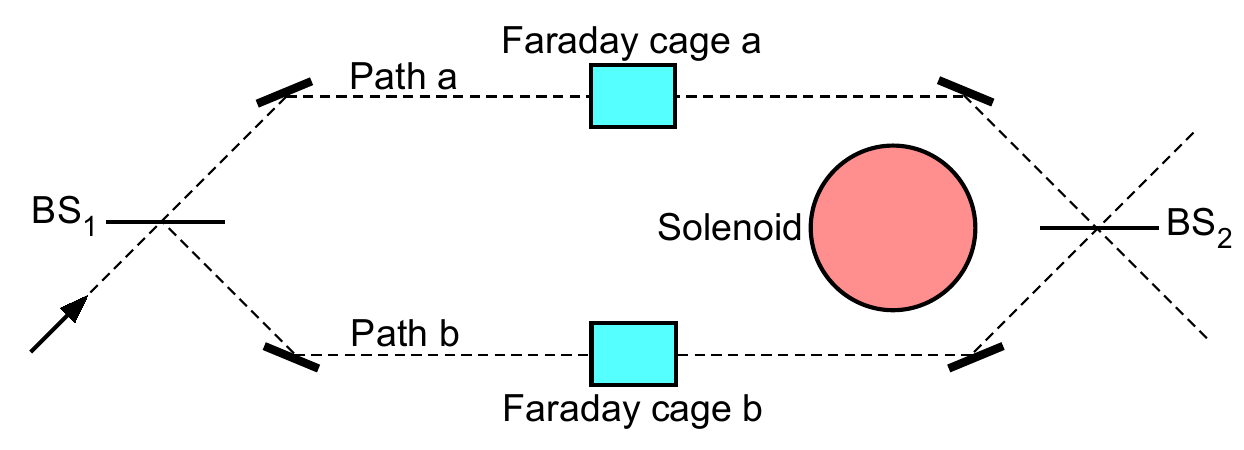}
  \caption{Mach-Zehnder interferometer for the implementation of different Aharonov-Bohm (AB) schemes. BS$_1$ and BS$_2$ are beam splitters. The quantum charged particle enters the interferometer by the indicated arrow. With the Faraday cages removed and the ``infinite'' solenoid with a magnetic flux $\Phi_0$, we have the magnetic AB effect \cite{ehrenberg49,aharonov59}. With the solenoid removed and the Faraday cages $a$ and $b$ subjected to different scalar potentials while the quantum charged particle is in a superposition state inside both cages, we have the electric AB effect \cite{aharonov59}. In the electrodynamic AB effect \cite{saldanha23}, the magnetic flux in the solenoid is varied while the quantum particle is in a superposition state inside both cages.}
\end{figure}

In the presence of controllable electromagnetic scalar and vector potentials $V$ and $\mb{A}$ in the interferometer, associated to electromagnetic fields that are not very intense, we include the following interaction term in the system Hamiltonian \cite{cohen2}:
\begin{equation}\label{HI}
	H_I = qV -\frac{q}{m}\mb{p}\cdot\mb{A},
\end{equation}
where $q$ and $m$ are the particle charge and mass, respectively, and $\mb{p}$ the average wave packet momentum.  In cases where each component of the wave function only propagates in regions with a null field, the term (\ref{HI}) modifies the previous wave functions by adding phases: $\Psi_i(\mb{r},t)=\Psi_i^0(\mb{r},t)e^{i\phi_i}$ with $i=\{a,b\}$ and $\phi_i=-\int_{t_0}^t H_I dt'/\hbar$, with $t_0$ being the time when the incident wave function is split by BS$_1$. Using Eq. (\ref{HI}), we can write
\begin{equation}\label{AB}
	\phi_{i}=-\frac{q}{\hbar}\int_{t_0}^t Vdt' +\frac{q}{\hbar}\int_{\mb{r}_0}^{\mb{r_i}}\mb{A}\cdot d\mb{x},
\end{equation}
where $\mb{p}/m$ was substituted by the average wave packet velocity $d\mb{x}/dt$, $\mb{r}_0$ represents the position where the incident wave packet is divided in BS$_1$, and the spatial integral is performed through the particle path.

In the scheme of Fig. 1, if the Faraday cages are absent (or are in a null potential) and the solenoid has a magnetic flux $\Phi_0$, we have the magnetic AB scheme \cite{ehrenberg49,aharonov59}. In this case, we have an AB phase difference due to Eq. (\ref{AB}) given by $q\Phi_0/\hbar$. If the solenoid is absent (or there is a null magnetic flux) and the scalar potentials in each Faraday cage vary while the particle is in a superposition state inside both cages, having the values $V_a$ (in path $a$) and $V_b$ (in path $b$) in this situation and being null while the quantum particle is outside the cages, we have the electric AB scheme \cite{aharonov59}. The AB phase difference due to Eq. (\ref{AB}) is then $(-q/\hbar)\int (V_a-V_b)dt$. In the electrodynamic AB effect \cite{saldanha23}, the magnetic flux in the solenoid varies while the quantum particle is in a superposition state inside both Faraday cages. The cages then shield the quantum particle from the induced electric field generated by the solenoid current variation. If, for instance, the solenoid magnetic flux is reduced to zero while the quantum particle is in a superposition state inside both cages in the scheme of Fig. 1, the final vector potential (in the Lorenz gauge) would be zero. Thus, only the interaction of the particle with the initial potential vector during its propagation from BS$_1$ to the Faraday cages would contribute for the AB phase difference, that would have a reduced value in relation to the magnetic AB effect where the solenoid magnetic flux does not change. A scalar potential difference between the cages could appear due to the electric charges that are induced in the cages to cancel their internal electric fields, but this contribution can be negligible depending on the system geometry (for instance, for very small Faraday cages). Interestingly, we may have a nonzero and gauge-independent AB phase due to Eq. (\ref{AB}) even if the solenoid is outside the interferometer \cite{saldanha23}.

\section{AB effect with a quantum electromagnetic field in the Lorenz gauge}

In previous works \cite{saldanha21a,saldanha21b}, we've used a second-order perturbative treatment to describe the AB effect with a quantum electromagnetic field in the Lorenz gauge. Here we obtain the same results in a different way, inspired in the recent work from Hayashi \cite{hayashi23}. First we obtain an exact solution for the electromagnetic ground energy considering the interaction of the quantum electromagnetic field with classical charges and currents. Then we include the interaction with the quantum particle using first-order perturbation theory, obtaining an expression for the AB phase. The advantage of this method is that it can be more easily adapted to compute the AB phase with the quantum electromagnetic potentials in other gauges. 

\subsection{Interaction of the quantum field with classical charges and currents}\label{subsec1}

The Hamiltonian for the free electromagnetic quantum field can be written as \cite{cohen}
\begin{equation}\label{H0}
	\hat{H}_0=\int d^3 k \sum_{\sigma}\left[ \hat{a}^\dag_{\sigma}(\mb{k})\hat{a}_{\sigma}(\mb{k})+\frac{1}{2} \right]\hbar\omega,
\end{equation}
with one quantum Harmonic oscillator for each field mode, defined by the wave vector $\mb{k}$ and polarization index $\sigma$ ($\omega=ck$ is the angular frequency). Hats are used in the field operators here. $\hat{a}_{\sigma}(\mb{k})$ and $\hat{a}^\dag_{\sigma}(\mb{k})$ are the annihilation and creation operators for the modes, that obey $[\hat{a}_{\sigma}(\mb{k}),\hat{a}^\dag_{\sigma'}(\mb{k}')]=\delta^3(\mb{k}-\mb{k}')\delta_{\sigma,\sigma'}$. In the traditional quantum optics treatments present in textbooks, constructed in the Coulomb gauge, the index $\sigma$ has two values for each wave vector, corresponding to two orthogonal transverse polarizations \cite{mandel}. But for a covariant treatment in the Lorenz gauge, the index $\sigma$ assumes 4 values: 0 for scalar photons, 1 and 2 for two transverse polarizations, and 3 for longitudinal photons \cite{cohen}. Scalar and longitudinal photons cannot carry energy or momentum between different regions of space, and the state of the free electromagnetic field never has such photons. But they are important to describe the electromagnetic interactions and they permit to write the interaction Hamiltonian of the field with charge and current densities $\rho$ and $\mb{J}$ in the covariant way
\begin{equation}\label{V}
\hat{H}_1=\int d^3r [\rho \hat{V} - \mb{J}\cdot\hat{\mb{A}}],
\end{equation}
where $\hat{V}$ and $\hat{\mb{A}}$ are the scalar and vector potential operators, given by
\begin{equation}\label{U}
	\hat{V}(\mb{r})=c\int d^3k \sqrt{\frac{\hbar}{2\varepsilon_0\omega(2\pi)^3}}\hat{a}_0(\mb{k})e^{i\mb{k}\cdot\mb{r}}-\mathrm{H.c.},
\end{equation}
\begin{equation}\label{A}
	\hat{\mb{A}}(\mb{r})=\int d^3k \sum_j \sqrt{\frac{\hbar}{2\varepsilon_0\omega(2\pi)^3}}\;\bs{\epsilon}_{\mb{k}j}\hat{a}_j(\mb{k})e^{i\mb{k}\cdot\mb{r}}+\mathrm{H.c.},
\end{equation}
with $j=\{1,2,3\}$ and $\bs{\epsilon}_{\mb{k}j}$ being a unitary polarization vector \cite{cohen}. The charge and current densities $\rho$ and $\mb{J}$ are to be considered the sources of the classical potentials in the schemes of Fig. 1. Note the minus sign present in the Hermitian conjugate term in Eq. (\ref{U}), which is important for the consistency of the theory. 

Now let us diagonalize the Hamiltonian $\hat{H}=\hat{H}_0+\hat{H}_1$. First let us define the quantities
\begin{equation}\label{l0}
	\lambda_0(\mb{k})= \frac{c}{\hbar\omega} \sqrt{\frac{\hbar}{2\varepsilon_0\omega(2\pi)^3}} \int d^3r \rho(\mb{r})e^{-i\mb{k}\cdot\mb{r}},
\end{equation}
\begin{equation}\label{ls}
	\lambda_j(\mb{k})= \frac{1}{\hbar\omega} \sqrt{\frac{\hbar}{2\varepsilon_0\omega(2\pi)^3}} \int d^3r [\mb{J}(\mb{r})\cdot\bs{\epsilon}_{\mb{k}j}]e^{-i\mb{k}\cdot\mb{r}},
\end{equation}
which are proportional to the spatial Fourier transforms of the charge density and of the current density components. By defining the operators
\begin{eqnarray}\nonumber\label{b}
	&&\hat{b}_0(\mb{k})=\hat{a}_0(\mb{k})-\lambda_0(\mb{k}),\;\;\hat{b}_0'^\dag(\mb{k})=\hat{a}_0^\dag(\mb{k})+\lambda_0^*(\mb{k}),\\
	&&\hat{b}_j(\mb{k})=\hat{a}_j(\mb{k})-\lambda_j(\mb{k}),\;\;\hat{b}_j'^\dag(\mb{k})=\hat{a}_j^\dag(\mb{k})-\lambda_j^*(\mb{k}),
\end{eqnarray}
 we can write
\begin{equation}\label{H}
	\hat{H}=\hat{H}_0+\hat{H}_I=\int d^3 k \sum_{\sigma}\left[ \hat{b}'^\dag_{\sigma}(\mb{k})\hat{b}_{\sigma}(\mb{k})+\frac{1}{2} \right]\hbar\omega+C,
\end{equation}
with
\begin{equation}\label{DE}
	C = \int d^3k\left[ |\lambda_0(\mb{k})|^2 -\sum_j|\lambda_j(\mb{k})|^2 \right]\hbar\omega
\end{equation}
being a constant. Since we have $[\hat{b}_{\sigma}(\mb{k}),\hat{b}'^\dag_{\sigma'}(\mb{k}')]=\delta^3(\mb{k}-\mb{k}')\delta_{\sigma,\sigma'}$, $\hat{H}$ from Eq. (\ref{H}) is a combination of quantum harmonic oscillators. The state of minimum energy, corresponding to the ground state $\ket{\tilde{0}}$ of the Hamiltonian (\ref{H}), obeys
\begin{equation}
	\hat{b}_\sigma(\mb{k})\ket{\tilde{0}}=\big[\hat{a}_\sigma(\mb{k})-\lambda_\sigma(\mb{k})\big]\ket{\tilde{0}}=0.
\end{equation}
So, the ground state $\ket{\tilde{0}}$ is a coherent state \cite{mandel} in the basis of the operators $\hat{a}_\sigma$, with 
\begin{equation}\label{a-coh}
	\hat{a}_\sigma(\mb{k})\ket{\tilde{0}}=\lambda_\sigma(\mb{k})\ket{\tilde{0}}.
\end{equation}
According to Eq. (\ref{b}), we also have
\begin{equation}\label{adag-coh}
	\bra{\tilde{0}}\hat{a}^\dag_0(\mb{k})=-\lambda^*_0(\mb{k})\bra{\tilde{0}},\;\;
	\bra{\tilde{0}}\hat{a}^\dag_j(\mb{k})=\lambda^*_j(\mb{k})\bra{\tilde{0}}.
\end{equation}

If the charge and current densities $\rho$ and $\mb{J}$ vary in an adiabatic way and the initial state for the quantum electromagnetic field is the ground state, it will always remain in the ground state $\ket{\tilde{0}}$ for the Hamiltonian of Eq. (\ref{H}).

\subsection{Interaction of the quantum field with a quantum charged particle and the AB phase}\label{subsec2}

Now let us introduce the interaction of the quantum electromagnetic field with a quantum charged particle in the nonrelativistic regime. The total Hamiltonian becomes $\hat{H}_T=\hat{H}+\hat{H}_2$, with $\hat{H}$ being the Hamiltonian treated in the previous subsection, with ground state $\ket{\tilde{0}}$, and 
 \begin{equation}\label{H2}
	\hat{H}_2 = q\hat{V} -\frac{q}{m}\mb{p}\cdot\hat{\mb{A}},
\end{equation}
with $\hat{V}$ and $\hat{\mb{A}}$ being the potential operators of Eqs. (\ref{U}) and (\ref{A}). As before, let us consider that the quantum particle wave function in each path of the interferometer $\Psi_i(\mb{r},t)$, with $i=\{a,b\}$, is known and has a reasonably well defined momentum, such that $\mb{p}$ in Eq. (\ref{H2}) can be considered the average particle momentum $\mb{p}_i(t)$ in each path. We also consider that the wave functions in each path have reasonably well defined positions, given by $\mb{r}_i(t)$.

With these considerations, treating the term $\hat{H}_2$ as a perturbation in the Hamiltonian $\hat{H}$, with the unperturbed state of the quantum electromagnetic field being $\ket{\tilde{0}}$, we obtain the first-order correction in the system energy
\begin{eqnarray}\nonumber
	\Delta E_i&\approx&\int d^3r|\Psi_i(r)|^2 \bra{\tilde{0}}\hat{H}_2(r)\ket{\tilde{0}}\\
	          &\approx& q  \bra{\tilde{0}}\hat{V}(\mb{r}_i)\ket{\tilde{0}} - \frac{q}{m} \mb{p}_i\cdot\bra{\tilde{0}}\hat{\mb{A}}(\mb{r}_i)\ket{\tilde{0}}.
\end{eqnarray}
Using Eqs. (\ref{U}), (\ref{A}),  (\ref{a-coh}), and (\ref{adag-coh}), we have
\begin{eqnarray}\nonumber
	\Delta E_i&\approx&\int d^3 r' \int d^3k \,q \frac{e^{i\mb{k}\cdot(\mb{r}-\mb{r}')}}{\varepsilon_0\omega^2(2\pi)^3}\times\\	&&\times 	\Big\{c^2\rho(\mb{r}')+\sum_j \left[\mb{J}(\mb{r}')\cdot\bs{\epsilon}_{\mb{k}j}\right] \left[\frac{\mb{p}_i}{m}\cdot\bs{\epsilon}_{\mb{k}j}\right]\Big\}.
\end{eqnarray}
Using the relations $\int d^3k {e^{i\mb{k}\cdot\mb{r}}}/{[(2\pi)^3k^2]} = {1}/{(4\pi|\mb{r}|)}$, $\omega=ck$, and $c=1/\sqrt{\mu_0\varepsilon_0}$, we obtain
\begin{equation}\label{DEf}
	\Delta E_i\approx q\mathcal{V}(\mb{r}_i)-\frac{q}{m}\mb{p}_i\cdot\mb{\mathcal{A}}(\mb{r}_i),
\end{equation}
where we defined the following effective scalar and vector potentials generated by the charge and current densities $\rho$ and $\mb{J}$:
\begin{equation}\label{UA1}
	{\mathcal{V}}(\mb{r})\equiv\bra{\tilde{0}}\hat{V}(\mb{r})\ket{\tilde{0}}=\int d^3r' \frac{\rho(\mb{r}')}{4\pi\varepsilon_0|\mb{r}-\mb{r}'|},
\end{equation}
\begin{equation}\label{UA2}
	\mb{\mathcal{A}}(\mb{r})\equiv\bra{\tilde{0}}\hat{\mb{A}}(\mb{r})\ket{\tilde{0}}=\int d^3r' \frac{\mu_0\mb{J}(\mb{r}')}{4\pi|\mb{r}-\mb{r}'|},
\end{equation}
which are equivalent to the classical potentials in the Lorenz gauge when we disregard the fields propagation time within the interferometer \cite{jackson}. 

Note that Eq. (\ref{DEf}) is equivalent to the expectation value of an effective Hamiltonian $\mathcal{H}$  for the quantum particle equivalent to Eq. (\ref{HI}) if we substitute $V$ and $\mb{A}$ by the effective potentials $\mathcal{V}$ and $\mb{\mathcal{A}}$ from Eqs. (\ref{UA1}) and (\ref{UA2}):
\begin{equation}\label{Hef}
	\mathcal{H}\equiv\bra{\tilde{0}}\hat{H}_2\ket{\tilde{0}} =q\mathcal{V} -\frac{q}{m}\mb{p}\cdot{\mb{\mathcal{A}}}.
\end{equation}
In this way, the AB phase difference from Eq. (\ref{AB}) follows directly. So, the energy of the ground state of the quantum electromagnetic field depends on the particle path, this behavior being the responsible for the appearance of the AB phase. 

The energy change of the ground state of the quantum electromagnetic field given by Eq. (\ref{DEf}) is the same as the one previously obtained with a second-order perturbative treatment \cite{saldanha21b}, which has an intuitive explanation as being associated to photon exchanges between the quantum particle and the other charge and current densities of the system. In this previous work \cite{saldanha21b}, it is also shown how the roles of the classical charge and current densities $\rho$ and $\mb{J}$ and the quantum charge and current densities $q|\Psi_i(r)|^2$ and $q|\Psi_i(r)|^2\mb{p}_i/m$ could be exchanged, with the quantum particle being the source of the effective potentials, that would them act on the classical charges and currents. This fact is closely related to Vaidman's work that describes the magnetic AB effect as the result of an influence of the field of the quantum charged particle in the source of the potential \cite{vaidman12}. 

\section{Gauge-invariance of the AB effect with a quantum electromagnetic field}

We can see that, by starting from a more fundamental treatment of the AB effect using a quantum electromagnetic field in the Lorenz gauge,  with the potential operators given by Eqs. (\ref{U}) and (\ref{A}), we obtain an effective Hamiltonian for the quantum particle given by Eq. (\ref{Hef}), with the classical effective potentials being also written in the Lorenz gauge, as in Eqs. (\ref{UA1}) and (\ref{UA2}). The effective Hamiltonian is equivalent to the energy change of the ground state of the quantum electromagnetic field due to the presence of the quantum particle, given by Eq. (\ref{DEf}), computed using perturbation theory. Now we see what changes if we use a quantum electromagnetic field in other gauges. 

A gauge change of the quantum electromagnetic field can be written as changes 
\begin{equation}\label{g-change}
	\hat{V}'=\hat{V}-\frac{\partial \hat{F}}{\partial t}\;,\;\; \hat{\mb{A}}'=\hat{\mb{A}} + \bs{\nabla}\hat{F}
\end{equation}
 in the scalar and vector potential operators from Eqs. (\ref{U}) and (\ref{A}), with 
\begin{eqnarray}\label{F}\nonumber
	\hat{F}(\mb{r},t)&=&\int d^3k \big[f_0(\mb{k},t)\hat{a}_0(\mb{k})e^{i\mb{k}\cdot\mb{r}}-\mathrm{H.c.}\big]+\\
	&&+\int d^3k \sum_j \big[f_j(\mb{k},t)\hat{a}_j(\mb{k})e^{i\mb{k}\cdot\mb{r}}+\mathrm{H.c.}\big]
\end{eqnarray}
being a scalar operator linear in the annihilation and creation operators and being Hermitian for the terms with $\hat{a}_j(\mb{k})$ and anti-Hermitian for the terms with $\hat{a}_0(\mb{k})$, in the same ways as the potential operators from Eqs. (\ref{U}) and (\ref{A}). Then, the total Hamiltonian is 
\begin{equation}\label{H-g}
	\hat{H}_T'=\hat{H}_0+\hat{H}_1+\hat{H}_2+\hat{H}_1'+\hat{H}_2',
\end{equation}
 with the first three terms on the right given by Eqs. (\ref{H0}), (\ref{V}), and (\ref{H2}) respectively, and  
\begin{equation}\label{VV}
	\hat{H}_1'=-\int d^3r \left[\rho \frac{\partial \hat{F}}{\partial t} + \mb{J}\cdot(\bs{\nabla}\hat{F})\right],
\end{equation}
\begin{equation}\label{VV2}
	\hat{H}_2'=-q \frac{\partial \hat{F}}{\partial t} - \frac{q}{m}\mb{p}\cdot(\bs{\nabla}\hat{F}).
\end{equation}
We will treat $\hat{H}_2$, and $\hat{H}_2'$ as perturbative terms in the total Hamiltonian $\hat{H}_T'$ from Eq. (\ref{H-g}).

An important distinction must be made at this stage, as discussed in the seminal work from Yang \cite{yang76}. For a time-independent Hamiltonian that describes a quantum charged particle  interacting with a classical electromagnetic field
\begin{equation}
	H= \frac{1}{2m}[\mb{p}-q{\mb{{A}}}]^2+qV,
\end{equation}
the Hamiltonian $H$ is equivalent to the system energy. In the adiabatic regime we are considering, even if the potentials change in time, as in the electric and electrodynamic AB effects, the AB phase can be computed considering constant potentials at each small time interval. But with a gauge change ${V}'={V}-\partial {F}/\partial t$ and ${\mb{A}}'={\mb{A}} + \bs{\nabla}{F}$, the novel Hamiltonian  
\begin{equation}\label{Ham}
	H'= \frac{1}{2m}[\mb{p}-q{\mb{{A}}}']^2+qV'
\end{equation}
in general is time-dependent and is not equivalent to the system energy. The system energy in the new gauge is given by \cite{yang76}
\begin{equation}\label{Ul}
	U'= H'+q\frac{\partial F}{\partial t}.
\end{equation}
The Hamiltonian $H'$ is the responsible for ruling the system evolution in the new gauge, while $U'$ represents the system energy. The Hamiltonians we consider in this work for the interaction of charges and currents (including the ones from quantum charged particles) with the electromagnetic field, as in Eqs. (\ref{HI}), (\ref{V}), and (\ref{H2}), are compatible with the one from Eq. (\ref{Ham}) when the terms with $q^2A^2$ are disregarded (they are important only for very intense fields). So, care must be taken with the energy operator in the new gauge. Considering $\hat{H}_0+\hat{H}_1+\hat{H}_1'$ as the unperturbed terms in the Hamiltonian of Eq. (\ref{H-g}), the energy eigenstate would not be the unperturbed Hamiltonian eigenstate. The term $q\partial F/\partial t$ from Eq. (\ref{Ul}) cancels the term with $\partial \hat{F}/\partial t$ of $\hat{H}_1'$ from Eq. (\ref{VV}) for the continuous charge distribution $\rho$. Also, using the relation $\mb{J}\cdot (\bs{\nabla} \hat{F})=\bs{\nabla}\cdot(\hat{F}\mb{J})-\hat{F}(\bs{\nabla}\cdot \mb{J})$ and the condition $\bs{\nabla}\cdot \mb{J}=0$ in the adiabatic regime we consider, we conclude that the term with $\mb{J}\cdot (\bs{\nabla} \hat{F})$ of $\hat{H}_1'$ from Eq. (\ref{VV}) is null. So, we conclude that the energy operator with the gauge change in the absence of the quantum particle, related to the unperturbed Hamiltonian $\hat{H}_0+\hat{H}_1+\hat{H}_1'$,  is the same energy operator as in the Lorenz gauge, given by $\hat{H}_0+\hat{H}_1$. On this way, the ground state for the quantum electromagnetic field in the absence of the quantum particle is the same that was computed in subsection \ref{subsec1}, given by the coherent state $\ket{\tilde{0}}$, that obeys Eq. (\ref{a-coh}). This makes sense, since an energy eigenstate should not depend on the chosen gauge.


We may write the novel effective Hamiltonian with the gauge change using Eqs. (\ref{g-change}), (\ref{F}), (\ref{a-coh}), (\ref{adag-coh}), (\ref{UA1}), and (\ref{UA2}). For the effective potentials, we obtain
\begin{equation}
	\mathcal{V}'\equiv\bra{\tilde{0}}\hat{V}\ket{\tilde{0}}-\bra{\tilde{0}}\frac{\partial \hat{F}}{\partial t}\ket{\tilde{0}}=\mathcal{V}-\frac{\partial \mathcal{F}}{\partial t},
\end{equation}
\begin{equation}
	\mathcal{A}'\equiv\bra{\tilde{0}}\hat{\mb{A}}\ket{\tilde{0}}+\bra{\tilde{0}}\bs{\nabla}\hat{F}\ket{\tilde{0}}=\mathcal{A}+ \bs{\nabla}\mathcal{F},
\end{equation}
 with $\mathcal{V}$ and $\mathcal{A}$ given by Eqs. (\ref{UA1}) and (\ref{UA2}), and 
\begin{equation}
	\mathcal{F}\equiv\bra{\tilde{0}}\hat{F}\ket{\tilde{0}}=\sum_\sigma\int d^3k f_\sigma(\mb{k},t)\lambda_\sigma(\mb{k})e^{i\mb{k}\cdot\mb{r}}+ \mathrm{c.c.}
\end{equation}
 The effective Hamiltonian in the novel gauge, obtained from Eqs. (\ref{H2}) and (\ref{VV2}), becomes
\begin{equation}\label{Hef-g}
	\mathcal{H}'\equiv\bra{\tilde{0}}[\hat{H}_2+\hat{H}_2']\ket{\tilde{0}} =q\mathcal{V}' -\frac{q}{m}\mb{p}\cdot{\mb{\mathcal{A}}}'.
\end{equation}
We thus see that the novel effective Hamiltonian when we perform a gauge change in the potential operators for the quantum electromagnetic field, initially in the Lorenz gauge, can be obtained from a gauge change in the effective potentials of Eqs. (\ref{UA1}) and (\ref{UA2}), initially in the Lorenz gauge. In this way, the AB phase difference from Eq. (\ref{AB}) follows directly, but now with the effective potentials in the novel gauge. Since the AB phase difference for a closed path with a classical treatment for the electromagnetic field is gauge-invariant, we have shown that this phase difference is also gauge-invariant with a quantum treatment for the electromagnetic field.

As previously discussed, the system energy in the novel gauge is different from the Hamiltonian, as shown in Eq. (\ref{Ul}). Only the term proportional to $\mb{p}\cdot(\bs{\nabla}\hat{F})$ of $H_2'$ from Eq. (\ref{VV2}) contributes for the energy, since the other one is canceled by the term $q{\partial F}/{\partial t}$ from Eq. (\ref{Ul}). Using first-order perturbation theory, the energy change of the ground state of the quantum electromagnetic field due to the presence of the quantum particle is
\begin{eqnarray}\label{DE-g}\nonumber
	\Delta E_i'&\approx&\int d^3r |\Psi_i(\mb{r})|^2\bra{\tilde{0}}\left[\hat{H}_2-\frac{q}{m}\mb{p}\cdot(\bs{\nabla}\hat{F})\right]\ket{\tilde{0}} \\
	&\approx&\Delta E_i-\frac{q}{m}\mb{p}_i\cdot[\bs{\nabla}\mathcal{F}(\mb{r}_i)],
\end{eqnarray}
with $\Delta E_i$ given by Eq. (\ref{DEf}). For a particular particle path, the AB phase can be written as
\begin{equation}
	\phi_{i}'=-\int_{t_0}^t dt' \frac{\Delta E_i'}{\hbar}=\phi_i+\frac{q}{\hbar}\int_{\mb{r}_0}^{\mb{r_i}}(\bs{\nabla}\mathcal{F})\cdot d\mb{x},
\end{equation}
where, as in Eq. (\ref{AB}), $\mb{p}/m$ was substituted by the average wave packet velocity $d\mb{x}/dt$, $\mb{r}_0$ represents the position where the incident wave packet is divided in BS$_1$, and the spatial integral is performed through the particle path. $\phi_i$ is given by Eq. (\ref{AB}) with the effective potentials $\mathcal{V}$ and $\mb{\mathcal{{A}}}$ from Eqs. (\ref{UA1}) and (\ref{UA2}) substituting the classical potentials $V$ and $\mb{A}$. Note that for two paths $a$ and $b$ that form a closed loop, with the same initial and final positions, we have $\phi_a'-\phi_b'=\phi_a-\phi_b$, a gauge-invariant AB phase difference.

\section{Discussion and Conclusion}

Note that, according to Eq. (\ref{DE-g}), the difference between the electromagnetic ground state energies for the quantum particle in the two interferometer paths may in general depend on the chosen gauge. Also, comparing Eqs. (\ref{Hef-g}) and (\ref{Hef}), we see that the quantum particle effective Hamiltonians also depend on the gauge. So, we conclude that the AB phase accumulated by a quantum particle in a particular path depend on the gauge used for the electromagnetic potential operators, as Hayashi showed for two specific gauges in Ref. \cite{hayashi23}.  Moreover, the effective Hamiltonian from Eq. (\ref{Hef-g}) is different from the energy change of Eq. (\ref{DE-g}), such that the predicted AB phase for a nonclosed path is different if we compute it using the effective Hamiltonian or using the particle influence on the electromagnetic ground energy.  So, for having a gauge-invariant AB phase, it is necessary to impose that the possible trajectories of the quantum particle in the interferometer form a closed path, with the initial and final positions of each possible path being the same. On this way, the resultant AB phase difference is gauge-independent, with the calculations using the effective Hamiltonian from Eq. (\ref{Hef-g}) agreeing with the calculations using the energy change of the electromagnetic ground state from Eq. (\ref{DE-g}). 

Marletto and Vedral recently proposed a way to measure the AB phase of a quantum charged particle without closing the interferometer coherently \cite{marletto20}, what could result in an intermediate AB phase. However, their proposal uses an ancillary particle. As discussed by Horvat \textit{et al.} \cite{horvat20}, when the phase accumulated by the ancillary particle is also taken into account, we obtain a gauge-invariant space-time loop integral involving the trajectories of both particles, which close the interferometer coherently and results in a full AB phase. We have also presented a proposal to measure an intermediate AB phase in Ref. \cite{saldanha21a}, which inspired the proposal of the electrodynamic AB effect \cite{saldanha23}. But in the detailed treatment of Ref. \cite{saldanha23} it is shown that in these cases the AB phase difference can be written as having contributions from a magnetic flux and from an electric flux in a spacetime surface whose boundaries are the possible particles trajectories in spacetime, being gauge-invariant. So, the situations considered in these works \cite{saldanha21a,saldanha23} correspond to closed paths with time-varying electromagnetic fields, resulting in gauge-invariant intermediate AB phase differences.

To conclude, we have presented a treatment for the AB effect using a quantum electromagnetic field. We have shown the gauge invariance of the obtained AB phase difference, by showing that a gauge change on the scalar and vector potential operators for the quantum field does not change the AB phase difference for closed paths. However, if it was possible to measure an AB phase difference for nonclosed paths, the resulting AB phase difference would not be gauge-invariant. So, we reinforce the conclusions from Horvat \textit{et al.} \cite{horvat20} and from Hayashi \cite{hayashi23} that it should not be possible to measure the AB phase difference in nonclosed paths.

\acknowledgements

This work was supported by the Brazilian agencies CNPq (Conselho Nacional de Desenvolvimento Científico e Tecnológico), CAPES (Coordenação de Aperfeiçoamento de Pessoal de Nível Superior), and FAPEMIG (Fundação de Amparo à Pesquisa do Estado de Minas Gerais).


%


\end{document}